\documentclass[conference]{IEEEtran}
\IEEEoverridecommandlockouts

\usepackage{caption}
\usepackage{subcaption}

\usepackage{cite}
\usepackage{amsmath,amssymb,amsfonts}
\usepackage{algorithmic}
\usepackage{booktabs}
\usepackage{colortbl}
\usepackage{graphicx}
\usepackage{textcomp}
\usepackage{xcolor}
\usepackage{float}
\usepackage{hyperref}
\def\BibTeX{{\rm B\kern-.05em{\sc i\kern-.025em b}\kern-.08em
    T\kern-.1667em\lower.7ex\hbox{E}\kern-.125emX}}
\begin{document}

\title{Biases in gendered citation practices: an exploratory study and some reflections on the Matthew and Matilda effects

}

\author{\IEEEauthorblockN{1\textsuperscript{st} Karolina Tchilinguirova}
\IEEEauthorblockA{\textit{Lassonde Engineering School} \\
\textit{York University}\\
Toronto, Canada \\
kdt@my.yorku.ca}
\and
\IEEEauthorblockN{2\textsuperscript{nd} Alvine B. Belle}
\IEEEauthorblockA{\textit{Lassonde Engineering School} \\
\textit{York University}\\
Toronto, Canada \\
alvine.belle@lassonde.yorku.ca}
\and
\IEEEauthorblockN{3\textsuperscript{rd} Gouled Mahamud}
\IEEEauthorblockA{\textit{Lassonde Engineering School} \\
\textit{York University}\\
Toronto, Canada \\
gouledm@my.yorku.ca}

}


\maketitle

\begin{abstract}
The number of citations of scientific articles has a huge impact on recommendations for funding allocations, recruitment decisions, promotion decisions and awards, just to name a few. 
Recent studies conducted in different scientific disciplines (e.g., physics and neuroscience) have concluded that researchers belonging to some socio-cultural groups (e.g., women, racialized people) are usually less cited than other researchers belonging to dominating groups. This is usually due to the presence of citation biases in reference lists. These citation biases towards researchers from some socio-cultural groups may inevitably cause unfairness and inaccuracy in the assessment of articles impact. These  citation biases may therefore translate to significant disparities in salaries, promotion, retention, grant funding, awards, collaborative opportunities, and publications. In this paper, we conduct –to the best of our knowledge – the first study aiming at analyzing gendered citation practices in the software engineering (SE) literature.  Our  study allows reflecting on citation practices adopted in the SE field and serves as a starting point for more robust empirical studies on the analyzed topic. Our results show that some efforts still need to be done to achieve fairness in citation practices in the SE field. Such efforts may notably consist in the inclusion of citation diversity statements in manuscripts submitted for publication in SE journals and conferences. Such efforts may also consist in redefining the power dynamics across scientific communities,  industry, and academia to foster scientific inclusion (e.g., fair representation of scientific contributions) in the SE field.

\end{abstract}


\section{Introduction}
Academic organizations usually assess each researcher based on the number of citations that the publications of that researcher yield \cite{b25}. Hence, the number of citations allows estimating the  impact that a given researcher has on a field \cite{b25}.  Thus, in many fields,  the higher the number of citations, the higher the likelihood of securing a job or a promotion in an academic organization, or obtaining grant funding as well as prestigious awards \cite{b27, b25}. In this regard, the reference lists of the papers we publish are powerful instruments since they have a major impact when it comes to determining who gets hired/promoted by an academic organization, who receives awards, who becomes famous in a field, etc.

A citation bias occurs when the authors of a given paper decide to include or not to include in their reference list a given reference based on considerations that are not related to the reference relevance and its quality \cite{b25}. Citation biases are detrimental to various socio-cultural groups, especially women and racialized researchers since their papers are usually under-cited \cite{b3, b12, b54}. For instance, in the most prolific countries, the papers in which women are the lead authors are usually less cited than the ones that have men as lead authors  \cite{b1}. Thus, citation biases may aggravate gender disparities especially since citations are key in the assessment of researchers \cite{b1}. That gender disparity is very pervasive and translates in academia in terms of research performance, the recruitment of faculty members, the number of citations researchers receive, grant applications, determination of the positions of authors in a paper, and the choice of academic awardees, just to name a few \cite{b2}. Besides, male researchers are more advantaged than female researchers when it comes to recruitment and promotion \cite{b2}. Furthermore, male researchers are usually more credited than female researchers when it comes to the contributions they make to research \cite{b2, b8}. This may give the false impression that female researchers are less capable than male researchers. This could notably be detrimental to the academic career of women \cite{b2} and to  innovation, creativity, performance, and productivity in various scientific and/or technological projects \cite{b18, b29, b30, b36, b39, b42}. That innovation is particularly key in the software engineering (SE) field since it drives technological advances, which influences various aspects of the society as a whole. Such aspects include the economy, the environment, as well as politics, just to name a few.

 The disparity of women in STEM (science, technology, engineering, and mathematics) disciplines is well-established \cite{b1, b3, b30, b31, b36, b38, b40, b41, b46, b55}. That disparity is well-pronounced in the software industry and has been increasingly explored (e.g., \cite{b35, b38, b40, b41, b42, b48, b49, b50}). Still, when it comes to research in particular, the gap in citations that researchers belonging to some socio-cultural groups may experience and that may stem from citation biases remains understudied  \cite{b3}. Furthermore, the awareness about that topic remains low  \cite{b3}. This makes it difficult to devise novel solutions to mitigate citation biases and make sure the scientific papers we publish sufficiently reflect the intellectual diversity of the existing scientific contributions.

 Several bibliometric studies have analyzed biases in citation practices  \cite{b2, b5, b6, b11, b12, b23, b1, b3, b9, b39}. These studies usually focus on STEM disciplines such as neuroscience, physics, economics, cognitive science, computer science, and biomedical sciences.  Such studies mostly focus on the analysis of citation biases toward women and to some extend towards racialized researchers. However, to the best of our knowledge, none of the existing work specifically applies to the SE field. This makes it challenging to assess the fairness of the citation practices adopted in that field. This also hampers the assessment of the impact that such citation practices may have on the recruitment, career, persistence, and prestige of researchers specialized in computing fields such as SE.

 To tackle that issue, we carry out a  study that relies on existing methods used in other disciplines (e.g., neuroscience, physics) to explore the fairness of gendered citation practices in the SE field.
The contributions of our paper are three-fold:
\begin{itemize}
    \item \textbf{Contribution 1}: we create a large dataset consisting of SE journals references and use it to analyze the fairness of gendered citation practices in the SE field.
    \item \textbf{Contribution 2}: we report our experience in using a set of statistical methods together with a set of APIs to support that citation analysis
    \item \textbf{Contribution 3}: we provide key insights on the fairness of citation practices in the SE field.
\end{itemize}

We further describe our work in the remainder of this paper. 

\section{Background and related work}
\label{section2}
\subsection{What is a Citation Bias?}
Citation bias involves selecting sources based on factors unrelated to their quality or relevance \cite{b25}. It is considered a Questionable Research Practice (QRP), which falls between responsible research conduct and research misconduct \cite{b15}. Citation bias can be deliberate or unintentional and stems from systematic and individual biases, including gender, race, ethnicity, geography, journal prestige, and favourable positive research results \cite{b25, b11}.
\newline 
\indent Citation bias significantly impacts both individuals in scientific communities and society. For researchers, citation metrics are crucial for career advancement, affecting jobs, promotions, grants, academic opportunities, awards, salaries, and collaboration \cite{b5, b11, b16, b25}. Additionally, citations reflect scientific inquiry, highlighting important questions and answers \cite{b5}. Thus, citation biases are detrimental to society as under-representing specific genders, races, nations, or research results can lead to several adverse outcomes. For instance, the lack of recognition for women's scientific achievements perpetuates stereotypes that discourage women and girls from entering the field, thus risking the erasure of women's contributions to scientific history \cite{b14}. Correspondingly, bias toward citing positive findings distorts the scientific processes, leading to incorrect conclusions and damaging science's reputation \cite{b15}. Furthermore, citation biases favouring resource-rich countries exclude contributions from emerging middle-income countries, limiting knowledge circulation and hindering innovation and sustained knowledge production \cite{b17}.

\subsection{Citation Biases in Literature}
The scope of studies on citation biases ranges from focusing on top journals within specific fields to broader, global analyses. The citation biases identified in these studies focus on gender, race, geographic origin, and positive results across disciplines such as neuroscience, physics, economics, and cardiology. In this section, we classify these studies based on the attributes they target.
\newline
\indent \textbf{Gender:}  Wu \cite{b44}  surveyed research focusing on the links between
gender and citations. Several studies (e.g., \cite{b1, b6, b9, b11, b14, b16, b19, b21, b39, b43, b45, b47})  highlighted persistent gender citation biases. These studies usually concluded that, despite an increase in women authorship, women-authored papers continue to receive fewer citations than men-authored papers. For instance, in economics, Ferber and Brun \cite{b16} found that male-only papers receive more citations than female-only or mixed-gender authored papers. Teich et al. \cite{b14} reported similar trends in physics, with male-authored papers being over-cited and women-authored papers under-cited, varying by sub field and proximity\cite{b14}. Larivière et al. \cite{b1} found that women in prominent author positions (sole, first and last) attracted fewer citations and consistently had more domestic portfolios with fewer international collaborations than men.  Dworkin et al. \cite{b11} showed that although women's representation increased in neuroscience, citation disparities persisted. Fulvio et al. \cite{b9} found that male-authored papers were still over-cited in cognitive neuroscience despite an increase in women-authored papers.
\newline
\indent
\textbf{Race:} The citation analysis Liu et al. \cite{b12} performed allowed them to conclude that papers authored by Black and Hispanic scientists are signiﬁcantly less cited compared to the ones authored by White and API (Asian and Paciﬁc Islander) scientists on similar topics. They observed that citational distortion accross several disciplines (e.g., engineering and computer science, physics and mathematics).
\newline
\indent
\textbf{Intersection of Race and Gender:} Bertolero et al. \cite{b5} extended the work of Dworkin et al. \cite{b11} to identify racial biases in neuroscience and explore the impact the intersection of gender and race has on citation practices. They found that authors of colour are increasingly under-cited, primarily by white authors,  black women being the most affected. They concluded  citation bias is most pronounced within racially segregated co-authorship networks, with white men receiving the most citations. 
\newline
\indent
\textbf{Geographic Origin:} Pasterkamp et al. \cite{b22} and Gomez et al. \cite{b17} identified biases favouring well-resourced countries. Pasterkamp et al. \cite{b22} found that in cardiology, citations primarily came from the USA and the recipient's own country, with a significant portion originating from the same institution. Gomez et al. \cite{b17} found that across 150 fields, research from highly active countries (e.g., USA, Western Europe, East Asia) is prioritized, while work from peripheral countries is overlooked. They concluded that the gap between core and peripheral countries is widening, especially in fields like physical sciences and engineering\cite{b17}.
\newline
\indent
\textbf{Positive Results:} Duyx et al. \cite{b15} determined that citation patterns favour positive results, particularly in biomedical sciences. They found that articles with significant results were cited 1.6 times more often, and articles which supported a researcher's hypothesis were cited 2.7 times more frequently. Positive articles receive about twice the citations of negative ones, influenced more by conclusions than data. Journal impact factor also affects citation rates as high-impact journals 
usually publish positive results and, thus, receive more citations \cite{b15}.
\newline
\indent
These studies underscore the systemic biases in citation practices, impacting the recognition and dissemination of scientific contributions across various demographics and regions. However, Ray et al. \cite{b25} found no peer-reviewed studies documenting racial or ethnic biases in citation practices within biology, medicine, chemistry, physics, or other natural sciences. Still, Liu et al. \cite{b12}'s recent study helps fill that gap.

\subsection{Citation Bias Detection Methods and Their Limitations}
Research on citation bias usually follows a similar workflow of data collection, data pre-processing, and statistical analysis. 
\newline
\indent\textbf{Data Collection:} there are three main types of approaches to data collection: 
\begin{enumerate}
    \item Collect metadata on papers from top journals; this approach is popular in research which focuses on a single field (e.g., \cite{b11, b5, b16, b14}).
    \item Collect metadata on a large set of papers from a variety of journals; this is an approach used for research interested in global citation patterns (e.g., \cite{b17, b1}).
    \item Collect papers based on a research topic;  Duyx et al. \cite{b15}, followed that approach that allows them to compile all papers which report on the association between article results and citation frequency.
\end{enumerate}
\indent\indent \textbf{Data Pre-processing:}  This usually allows labelling authors and authors cited within each paper  based on some interest criteria i.e attributes. Studies on gender bias used databases such as the US Census, Wikipedia, and Social Security Administration to match names to gender \cite{b16, b14, b1, b11, b5, b9}. Bertolero et al. \cite{b5} assigned race using probabilistic databases and neural networks trained on American Voter and Census data. Geographic bias studies such as the one of Gomez et al. \cite{b17} extracted author location information.  Duyx et al. \cite{b15} extracted information from the papers content, which included information such as the number of positive articles, the number of negative articles, the number of citations to positive articles, and the number of citations to negative articles. Note that, at this stage, the removal of self-citations from reference lists is common practice \cite{b11, b5, b16, b14, b9, b22}.
\newline
\indent \textbf{Statistical Analysis:} Several studies (e.g., \cite{b5, b6, b9, b14}) on citation analysis used/adapted Dworkin et al. \cite{b11}'s statistical approach. Thus, most of them created a model to determine expected citation rate by gender/race and compared them to actual rates to identify any citation biases. Gomez et al. \cite{b17} analyzed global citation behaviour using a three-layer network (citation, text similarity, and distortion) to understand how countries cite each other and the similarities between their research topics. Duyx et al. \cite{b15} examined the relationship between citation frequency and the statistical significance, direction, hypothesis conformity, and authors' conclusions of the article results.
\newline
\indent \textbf{Limitations:}  Studies which extract data from top journals must be careful with generalizing bias within the field as a whole \cite{b11, b5, b14}. Dworkin et al. \cite{b11} state that they did not account for authors' institutional prestige, potentially introducing bias due to gender imbalance in hiring and prestige-based citation behaviour. Gender-based papers such as  Dworkin et al. \cite{b11} caution that gender determination methods are binary, excluding intersex, transgender, and non-binary identities. Bertolero et al. \cite{b5} investigated race bias, which is limited by probabilistic racial and ethnic identity analyses based on limited racial categories and outdated census data that may inaccurately assign categories. Furthermore, Gomez et al. \cite{b17} acknowledge that their citational lensing framework is limited in accuracy as the textual similarity between countries is inherently noisy, which impacts comparisons with citation data. Duyx et al. \cite{b15}'s meta-study  on result-biased citations acknowledges the need for caution in making broad conclusions about citation bias across all sciences due to the high heterogeneity in their analyses. 

\subsection{Citation Biases Mitigation Methods and their Limitations}
The literature proposed several solutions to mitigate citation biases. For instance, Larivière et al. \cite{b1, b26} detected global gender bias against women and recommended fostering international collaborations programs for women researchers to reduce disparities in research output and impact. Additionally, they state that policymakers should consider the diverse contexts in which science is performed and identify mechanisms that perpetuate gender inequality. 

Teich et al. \cite{b14} found that women are cited more in longer reference lists, thus noting that removing length limits on reference lists could promote gender equality.  However, this should not imply that man-authored papers are more valuable. Bertolero et al.\cite{b5} recognized racial citation bias against authors of colour, and, along with Dworkin et al. \cite{b11},  they advocate for education, transparency, personal responsibility, and ally-ship within research communities. Additionally, they emphasize that non-minority male scholars must recognize and address the challenges they create and perpetuate.

 Duyx et al. \cite{b15} stated that journals should include declarations about the representativeness of the cited literature, akin to those for funding and author contributions. This  could help raise awareness on selective citation practices and mitigate result-based citation bias. Along the same lines, several authors such as Rowson et al. \cite{b10},  Dworkin et al. \cite{b11}, Teich et al. \cite{b14}, and Ray et al. \cite{b25} advocate for the inclusion of Citation Diversity Statements (CDSs) in papers. 
 A CDS is a paragraph added before the reference list of a paper \cite{b25}.
 It shows a commitment to citation equity \cite{b25}. Full CDSs provide a percentage breakdown of citations, explain assessment methods and limitations, and highlight citation diversity  \cite{b10, b25}. 

Ray et al. \cite{b25} highlight several challenges with CDSs, including the complexity of addressing various aspects of diversity, accurately identifying them, and assessing whether citation diversity reflects a discipline's diversity. They emphasize that citations should be for scholarly purposes to avoid unethical practices, while recognizing the risk of tokenism, where authors may feel compelled to diversify citations to appear supportive of DEI (diversity, equity,
inclusion) efforts.

\section{Methodology}
\label{section3}
In this paper, we focus on citation bias detection. The goal of this paper is therefore to study the citation practices in the SE field. More specifically, our study aims at determining if these citation practices are biased towards a  popular  attribute: the gender of authors. To perform our study, we first create a dataset (i.e. a reference list) consisting of references of papers published in the top SE journals. The  analysis of that dataset then allows us to determine if that dataset exhibits some citations biases toward female researchers. To support that analysis, we investigate three research questions (RQs):
\newline
\textbf{(RQ1):} What is the gender distribution in
SE journals authorship?
\newline
\textbf{(RQ2):} What is the temporal evolution of citation trends in the SE journals? 
\newline
\textbf{(RQ3):} Are there some citation biases toward female authors in papers published
in SE journals?

\subsection{Overview of our methodology}
\begin{figure}
    \centering
    \includegraphics[width=0.3\textwidth]{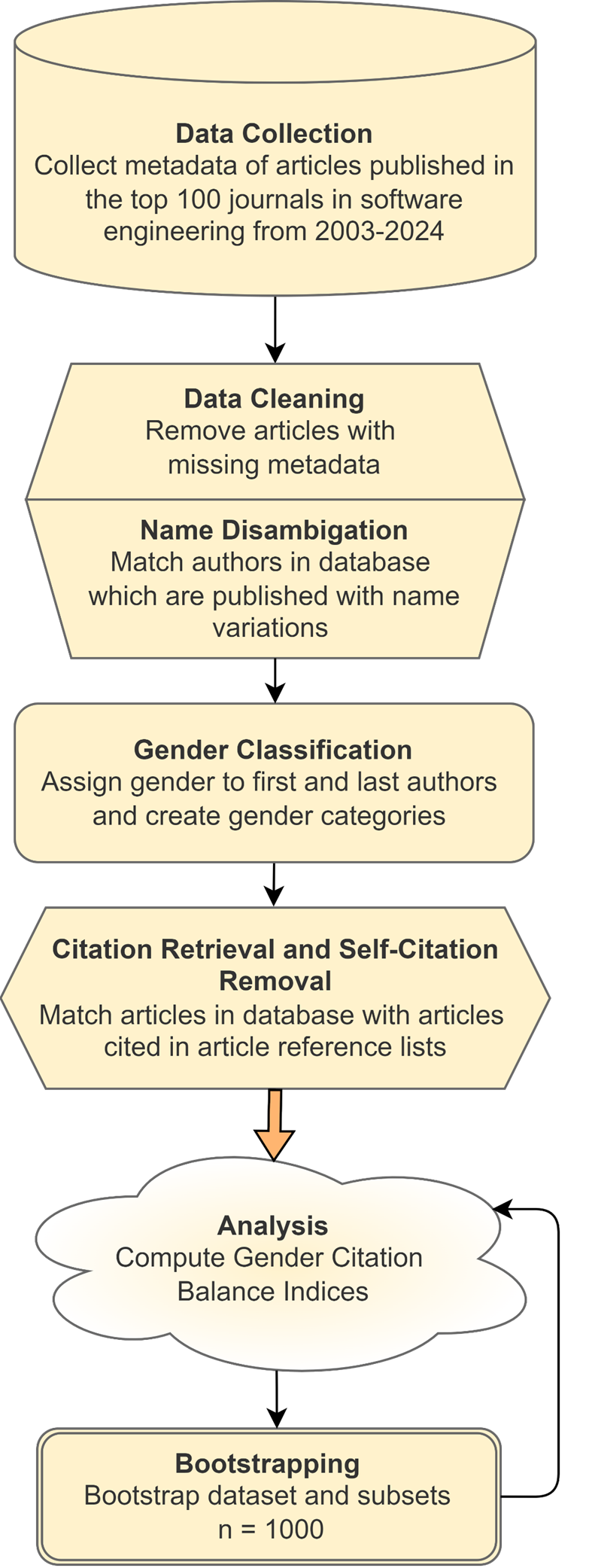}
    \caption{Methodology workflow}
    \label{fig:methodology_workflow}
\end{figure}

To perform our study, we adapted the citation bias detection methods that Dworkin et al. \cite{b11} as well as Fulvio et. al \cite{b9} proposed. Figure \ref{fig:methodology_workflow}  illustrates the six steps we followed to complete our study. 
We perform Step 1 through Step 4 using 
 Dworkin's \textit{R} code  \footnote{\href{https://github.com/jdwor/gendercitation}{Link to Dworkin et al.'s code}.}. We complete  Steps 5 and  6 using our own Python scripts, but drawing inspiration from Fulvio et al. \cite{b9}'s approach. We further describe our methodology below.

\subsection{Step 1: Data Collection }
We collected metadata from the top 100 software engineering journals \footnote{\href{https://anonymous.4open.science/r/ICSE_2025_SEIS_submission_codebase/journal_list.csv}{Link to the list of 100 journals}} as ranked on Web of Science (WoS) by Eigen- factor scores. We followed Dworkin et al. \cite{b11}’s data retrieval method which consisted of manually querying journals through \textit{Web Of Science.com}. We searched by publication title and specified one title at a time. We filtered the results to only include publication years from 2003 to 2024 and document types of Article, Review Article and Proceeding Paper. We manually exported the results using the WoS exporting tool, where we specified the record content as ”\textit{Full Record and Cited References}” and exported the maximum record count of 500 to Plain Text File. Thus, the data was downloaded 500 records at a time and one journal at time. The downloaded Plain Text File included all available metadata for each article, which included but not limited to fields such as article title, abstract, funding information, journal information and number of citations. The metadata of interest for this project includes publication date, DOI (Digital Object Identifier), author list and reference list of each article. This step allowed us to collect the metadata of 166,657 research articles, review articles and processing papers published in 100 SE journals from January 2003 to August 2024. From the 166,657 articles collected, 4996 articles had missing DOI's, thus we excluded them from the dataset. This reduced the dataset to 161,661 articles.

We stored the Plain Text Files in folders by journal and imported into Dworkin's \textit{R} codebase for processing. We stored the post processed dataset in  MongoDB --a document-oriented database -- which we use later for the statistical analysis. Our raw data and processed dataset are available online  \footnote{\href{https://anonymous.4open.science/r/ICSE_2025_SEIS_submission_codebase/README.md}{Link to the Raw data, see /R-processing/wos-data}} \footnote{\href{https://anonymous.4open.science/r/ICSE_2025_SEIS_submission_codebase/R-processing/article_data_jsons/article_data_part_1.json}{Link to the MongoDB dataset exported as a set of JSON files}}.

\subsection{Step 2: Name Processing}
Author names serve two important roles in our analysis: author gender assignment and self-citation removal. Still, name data is inconsistent as it may be incomplete such is the case where only initials are included in the metadata or authors  published under different name variations.  Thus, ensuring complete names will minimize missing data and increase successful gender assignment. Effectively linking author name variations together reduces the number of missed self-citations. 

We use Dworkin et al. \cite{b11}’s \textit{R}  code to process the metadata we collected in Step 1. In their code, Dworkin et al. \cite{b11} cross-reference any initial only names as provided by WoS with CrossRef by querying the CrossRef API with the article DOI. If the requests return a full name, then the WoS name is replaced. Secondly, Dworkin et al. implemented a name disambiguation algorithm which matches name variants to connect papers authored by the same person, but are published under different names.  This algorithm functions by matching all entries which share the first name initial and/or middle initial and last name are the same. Since, last names are matched first, followed by matching first and/or middle initials, it is crucial to ensure all last names are complete. 
In Dworkin et al. \cite{b11}’s words “\textit{For example, if an entry listed an author as R. J. Dolan, and we found matches under Ray J. Dolan and Raymond J. Dolan, we would replace the R. J. Dolan entry with the more common completed variant. If, instead, we found matches under Ray J. Dolan and Rebecca J. Dolan, we would not assign a name to the original R. J. Dolan entry.}” (pg. 11).  

Dworkin et al. \cite{b11} noted in their database all last names where complete in the initial WoS metadata. However, these was not the case for our dataset, therefore, we implemented an additional \textit{R}  script which removed any articles which has either initial or missing last names. In total, we removed 853 articles which had incomplete last names. Thus, this  reduced our dataset size to 160,808 articles.
The Secure Open Enterprise Master Patient Index is a scientific record linkage tool which contains a database of common nicknames. We used it to connect name variants \cite{b21}. Any names where the full names cannot be matched or retrieved with CrossRef will not be assigned a gender and is excluded from the analysis. In this step, CrossRef replaced a total of 19,164 names. Besides, 14,474 name variations matches were found.

\subsection{Step 3: Gender Assignment}
To assign gender to the authors in our processed data, we also rely on Dworkin et al. \cite{b11}'s code. Like in the literature (e.g., \cite{b5, b6, b9, b11, b14}), our analysis  only considers the gender of first and last authors of articles and references. This assignment allows organizing each dataset into four gender-based categories: 1) \textbf{MM} (first author is a man and last author is a man); 2)  \textbf{MW} (first author is a man and last author is a woman);  \textbf{WM} (first author is a woman and last author is a man); and 4)  \textbf{WW} (first author is a woman and last author is a woman). As in the literature (e.g., \cite{b11}), articles with a sole author are grouped into either the WW or MM gender category. Note that, for the sake of the citation analysis, we will introduce another category called \textbf{W$\cup$W}  (i.e. Woman or Woman) \cite{b11}. That category is the union of the following categories: WW, MW and WM. Thus, the W$\cup$W category refers to papers in which a female author is involved as the lead author, as the last author, or both. 

The gender assignment process Dworkin et al. \cite{b11}'s code supports consists in performing two rounds of gender assignment from two different databases. 
\begin{itemize}
    \item \textbf{First round of gender assignment}: in this round, the Social Security Administration (SSA) database as implemented in the \textit{gender} R package is used to assign genders to authors whose first names are available in the processed dataset.  This SSA database provides the proportion of male-female infants assigned to a name in the US from 1932 to 2012. If a given author's name yields a proportion greater or equal to \textbf{0.7}, then a gender is assigned to the corresponding author. Noteworthy, picking 0.7 as a threshold  is common in the literature (e.g., \cite{b9, b11, b43}).
    \item \textbf{Second round of gender assignment}: in this round, the paid service \textit{Gender-API} is used to fill in the gender data for any names which were either missing in the SSA database or returned a proportion between 0.3-0.7. Gender-API  has a database of 6,196,452 unique names from 191 countries which are sourced from both publicly available data and government data. Gender-API returns a probability value of the name gender which is based on the number of records for that name in the database \cite{b53}. The same threshold of \textbf{0.7} is applied to names assigned gender with Gender-API. 
\end{itemize}

Authors whose gender is not assigned are labelled as "U" for \textit{Unknown} and  will be excluded from our dataset. Reasons a gender could not be assigned include: only initial available, name not found in either database or gender probability below threshold. Completing this step allowed us to assign gender for both first and last authors of 66,63\% of the papers. Thus, 107,152 papers  had authors with fully assigned genders while 53,656 articles had at least one author  with \textit{Unknown} gender.

\subsection{Step 4: Citation Processing} Following Dworkin  et al. \cite{b11}'s methodology for our analysis, we do not consider all articles cited in the metadata reference lists, instead we only consider citations of articles which are in our dataset. Thus, we must identify any articles present in the metadata references lists that also exists in our dataset; all other referenced articles will be removed from the reference list. We explain this process below.

The reference lists of each article is supplied by the metadata from WoS. The reference list is provided as a string which is processed by Dworkin’s code. The string is processed to extract DOIs in the reference list string.  These extracted DOIs are then matched to any  DOIs in our database and matches are stored as indices which point to location of the associated article. This pointer supplies the author and gender data required for the analysis. 

As we mention in Section \ref{section2}, the removal of self-citations from reference lists is a well-established practice in citation analysis (e.g., \cite{b11, b5, b16, b14, b9, b22}). It allows mitigating the impact gender differences may have on self-citation patterns \cite{b9}. Dworkin et al. \cite{b11} define a self-citation as any paper listed in a reference list where the first or last author of that paper is also the first or last author of the citing paper. Dworkin et al.  acknowledge that this is a restrictive definition. However, due to the nature of this analysis, since cited genders will be based on assignments already present in our database, this is the only way we can guarantee  the authors are present in our database. We rely on Dworkin et al. \cite{b11}’s code to identify potential self-citations. This process consists of collecting a list of all first authors and a list of all last authors. These names are then matched to any papers where the authors name appear in that list. Each paper stores a list of indices which point to the papers in our database which have the same author. These indices represent potential self-citations.

We export the processed \textit{R}  Data to a MongoDB database. The processed dataset has now been gender labelled and indexed with pointers that connect cited papers, to the citing paper as well as papers authored by the same person.
This section of the Methodology workflow is where we deviate away from Dworkin et al. \cite{b11}’s statistical analysis that leverages a Generalized Additive Model (GAM), and adapt an analysis similar to Fulvio et al. \cite{b9}’s. 
This allows ensuring the citation processing step is suitable for the  journals we target. Hence, in accordance with Fulvio et al. \cite{b9}, we rely on our own Python script to automatically remove the self-citations we identified using the code of Dworkin et al. \cite{b11}. 
Thus, we consider that each article in the processed dataset includes a list of cited article indices and a list of article indices which are authored by the same person. If an article index appears in both lists, our script removes it from the cited list as it means it is a self citation. Besides, when assigning gender categories to cited articles, we use the list of cited article indices for each article in our database to create a new list which stores gender categories instead of indices.  

Since, the authors which were not able to be assigned gender were simply labeled as \textit{Unknown} but not immediately removed from the dataset in Step 3, the citation processing was conducted on the processed dataset (160,808 articles). In total 567,374 citations were retrieved and 38,243 of those were removed as they were self-citations. 
However, these citations include citing and cited articles which contain authors without assigned genders. Thus, when we filter those out of the dataset, we obtained 107,152 articles and 280,217 citations of which 25,561 were removed as self-citations. 
The citation data was very sparse up until 2009, with only 3,482 citations in total for those five years (i.e. 1.37 \% of all citations). Thus, in accordance with the literature (e.g., \cite{b6, b9, b11}), we decided at this stage to conduct our citation analysis for articles published from January 2009- August 2024. This will allow us in Step 5 to focus on citations
that yield a more robust statistical citation analysis. In the current Step (i.e. Step 4), when filtering out articles published from 2003-2008, we ended up with 96,282 articles and 276,187 citations of which 25,013 are self-citations. Thus, our final citation analysis dataset consists of 96,282 articles with 251,174 citations.

\subsection{Step 5: Statistical citation analysis } 
To simplify our citation analysis as  in the literature (e.g., \cite{b5, b6, b9, b11, b14, b47}), we only focus on the first and last authors of each paper available in our dataset. The last author is usually the most senior author in some fields \cite{b6, b47}.  To support the citation analysis, Fulvio et al. \cite{b9} rely on an index called the \textbf{Gender Citation Balance Index}. The latter allows comparing the amount of citation each gender category receives with respect to the existing distribution of gender categories in SE literature. Equation \ref{equation:equation1} formalizes that index. To perform our citation analysis, we apply that equation separately to each gender category.  A positive index yields a gender category that is more cited than expected (i.e. positive bias) \cite{b9}. A negative index yields a gender category that is less cited than expected (i.e. negative bias) \cite{b9}. 
\begin{equation}
\begin{aligned}
&\text{Gender Citation Balance Index} \\
&\quad= \frac{\text{observed proportion} - \text{expected proportion}}{\text{expected proportion}}
\end{aligned}
\label{equation:equation1}
\end{equation}


\subsubsection{Calculating expected proportion}
The expected proportion represents the percentage of papers in our dataset which are authored by a specified gender category. This is known as our expected proportion, because if there is no biases in the citations of papers, we should expect our proportion of citations to be the same as our proportion of authorship.  Equation \ref{equation:equation2} shows how we compute the expected proportion.
\begin{equation}
\begin{aligned}
&\text{Expected proportion of gender category} \\
&\quad= \frac{\sum \text{(papers authored by gender category)}}{\sum \text{(total papers)}} \times 100
\end{aligned}
\label{equation:equation2}
\end{equation}

Recall from Step 4 that we obtained a total of 96,282 papers. Table \ref{tab:expected} reports the corresponding papers breakdown per gender category. We use these papers to derive the expected proportions per gender categories and to reason about the authorship trends in the SE literature.

\begin{table}[H]
\caption{Paper breakdown for each gender category }
\centering
\begin{tabular}{|c|c|}
\hline
Gender Category & \#Papers \\
\hline
MM (Man first author – Man last author) & 	
69,239  \\
MW (Man first author – Woman last author) & 9,287  \\
WM (Woman first author – Man last author) & 13,529  \\
WW (Woman first author – Woman last author) & 4,227 \\
\hline
\end{tabular}
\label{tab:expected}
\end{table}

\subsubsection{Calculating observed proportion}
The observed proportion (see equation \ref{equation:equation3}) represents the percentage of citations which are authored by a specific gender category. It is known as our observed proportion, because it reflects the distribution of gender as it appears among citations in our dataset. 
\begin{equation}
\begin{aligned}
&\text{Observed proportion of gender category} \\
&\quad= \frac{\sum \text{(cited papers authored by gender category)}}{\sum \text{(total citations)}} \times 100
\end{aligned}
\label{equation:equation3}
\end{equation}

Recall from Step 4 that we obtained a total of 251,174 citations. Table \ref{tab:observed} reports the citation breakdown per gender category.  We use these citations to derive the observed citation proportions per gender categories and to reason about the citation trends in the SE literature.

\begin{table}[H]
\caption{Citation breakdown for each gender category}
\centering
\begin{tabular}{|c|c|}
\hline
Gender Category & \#Citations \\
\hline
MM (Man first author – Man last author) & 192,425 \\
MW (Man first author – Woman last author) & 21,861  \\
WM (Woman first author – Man last author) & 28,381  \\
WW (Woman first author – Woman last author) & 8,507  \\
\hline
\end{tabular}
\label{tab:observed}
\end{table}

\subsection{Step 6: bootstrapping}
Like Fulvio et al. \cite{b9}, we bootstrapped the 95\% confidence interval for each gender category using 1000 iterations of random sampling with replacement from the total number of papers obtained in Step 4 i.e.  96,282 papers. Thus, for each iteration, we computed the Gender Citation Balance Index for each category, considering: 1) the 2.5th percentile as the lower bound of the confidence interval; and 2) the 97.5th percentile as the upper bound of the confidence interval.

We performed these statistical steps  once for the entire dataset, then again for each set of papers (i.e. subset of the dataset) associated with a gender category. This allows analyzing the citing behaviour of the following gender categories: MM, WW, and WuW. When calculating the Gender Citation Balance Indices for the subsets, the expected proportion remains the same from the full dataset. Additionally, each subset is filtered out from the full dataset and bootstrapped individually (i.e. not subdivided from the already bootstrapped dataset).


\section{Result analysis}
\label{section4}
We relied on our methodology to perform citation analysis. In the remainder of this section, we further discuss the results of that analysis in the light of our three research questions.

\subsection{RQ1: gender distribution in
SE journals authorship}
\begin{figure*}
    \centering
    \includegraphics[width=0.78\linewidth]{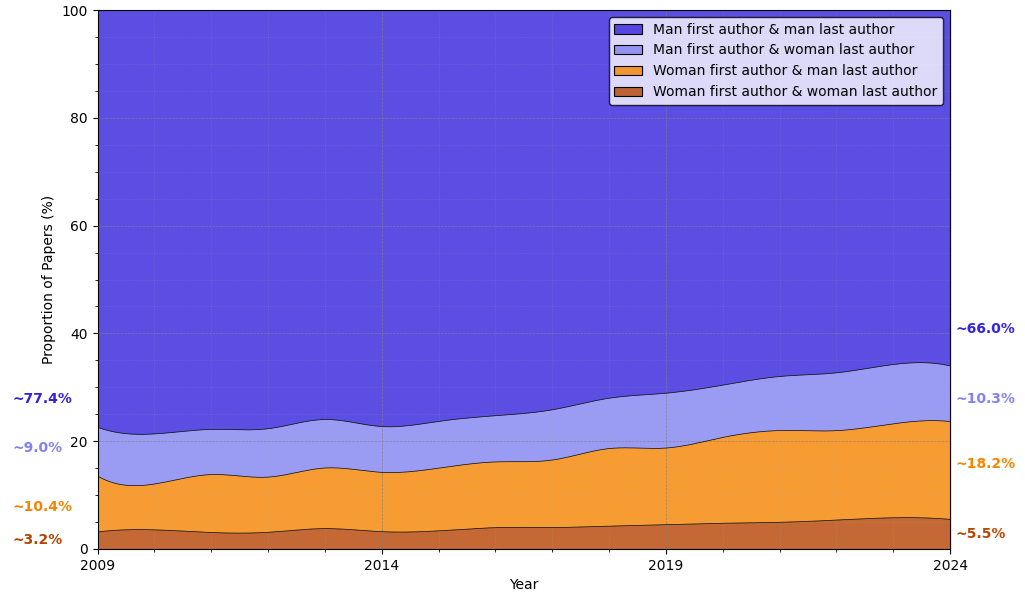}
    \caption{Expected proportions: authorship trends in the software engineering literature (2009 - 2024) }
    \label{fig:authorship_trends}
\end{figure*}

We completed Steps 1 to 4 of our methodology. This allowed us to use the resulting 96,282 papers to plot the authorship trends in the SE literature between 2009 and 2024. Figure  \ref{fig:authorship_trends} shows that chart. Hence, Figure \ref{fig:authorship_trends}  illustrates the expected proportions of papers authored in each gender category. That Figure shows that, in the SE literature, the MM category dominates the authorship while the WW category is underrepresented. Still, over time, more papers have been authored by authors belonging to the WW category. Hence, between 2009 and 2024, the WW authorship has increased by nearly 2.3\% in the SE field. But, as of August 2024, the authorship in the SE field remains little diversified since the MM authorship still represents two-thirds of the authorship while the WW authorship only represents 5.5\% of the authorship.


\smallskip
\noindent\fbox{%
    \parbox{\linewidth}{%
    \smallskip
       Our results indicate that, in the SE field, men write most of the scientific papers. The female authorship, in spite of its increase,  still remains significantly underrepresented.

    }}

\subsection{RQ2:  temporal evolution of citations 
trends in SE journals}
\begin{figure*}
    \centering
    \includegraphics[width=0.78\linewidth]{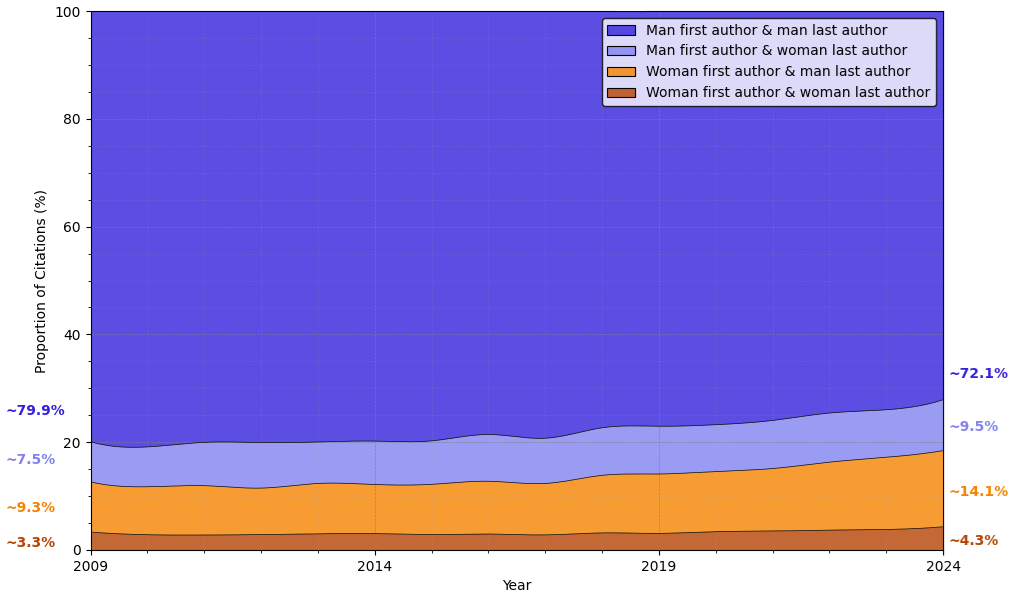}
    \caption{Observed proportions: temporal evolution of citation trends in the software engineering literature  (2009 - 2024)}
    \label{fig:citation_trends}
\end{figure*}

Recall from above that we obtained 251,174 citations when completing Step 4 of our methodology. We used these citations to plot the temporal evolution of citations trends in the SE literature. Figure  \ref{fig:citation_trends} illustrates the corresponding chart. Thus, that Figure shows the observed citation proportions for each gender category from 2009 to 2024.  Interestingly, when analyzing that chart, we can notice that diversity in the SE literature seems to have slightly increased over time, with female-led papers (i.e. papers in the WW category) being increasingly more cited as time passes. But that increase has been rather marginal throughout the years since the citation proportion of the WW category has only increased by 1\% from 2009 to 2024, while, as Figure \ref{fig:authorship_trends} shows, the proportion of female authors has nearly doubled during the same period. Besides, as  Figure  \ref{fig:citation_trends} shows, the most cited category is the MM category, while the least cited category remains the WW category. Thus, time has not sufficiently witnessed fairness in gendered citation practices.

\smallskip
\noindent\fbox{%
    \parbox{\linewidth}{%
    \smallskip
       Our results show that, in the software engineering literature, male-led papers are significantly more cited than women-led papers. That trend has been consolidated over time.}

    }
    
\subsection{RQ3: citation biases in papers published in SE journals}
We completed Steps 1 to 6 of our methodology and specifically relied on equation \ref{equation:equation1} (see Step 5)
to  compute the Gender Citation Balance Index for each of the gender categories. Figure  \ref{fig:citation_rate_full} depicts the corresponding results, with error
bars associated with bootstrapped
95\% confidence intervals (Step 6). Thus, that Figure depicts the biases in gendered citation practices in the SE literature from 2009 to 2024. As shown on that Figure, the MM category is over-cited i.e. positively biased. Contrariwise, the MW, WM and WW categories are significantly under-cited (i.e. negatively biased), the WW category being the most under-cited category. Gender-wise, several citation bias patterns could explain these citation bias results. Our analysis allowed us to identify three of such patterns: 

\begin{figure}
     \centering
     \begin{subfigure}[b]{0.5\textwidth}
         \centering
         \includegraphics[width=0.9\linewidth]{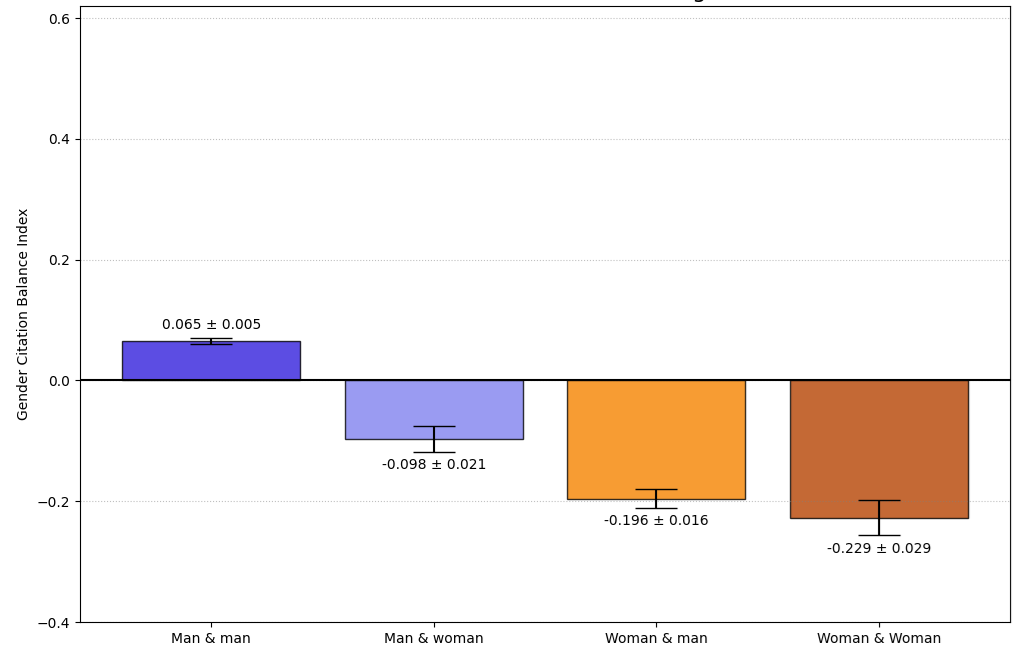}
    \caption{Biases in gendered citation practices}
    \label{fig:citation_rate_full}
     \end{subfigure}
     \hfill
     \begin{subfigure}[b]{0.5\textwidth}
         \centering
         \includegraphics[width=0.85\linewidth]{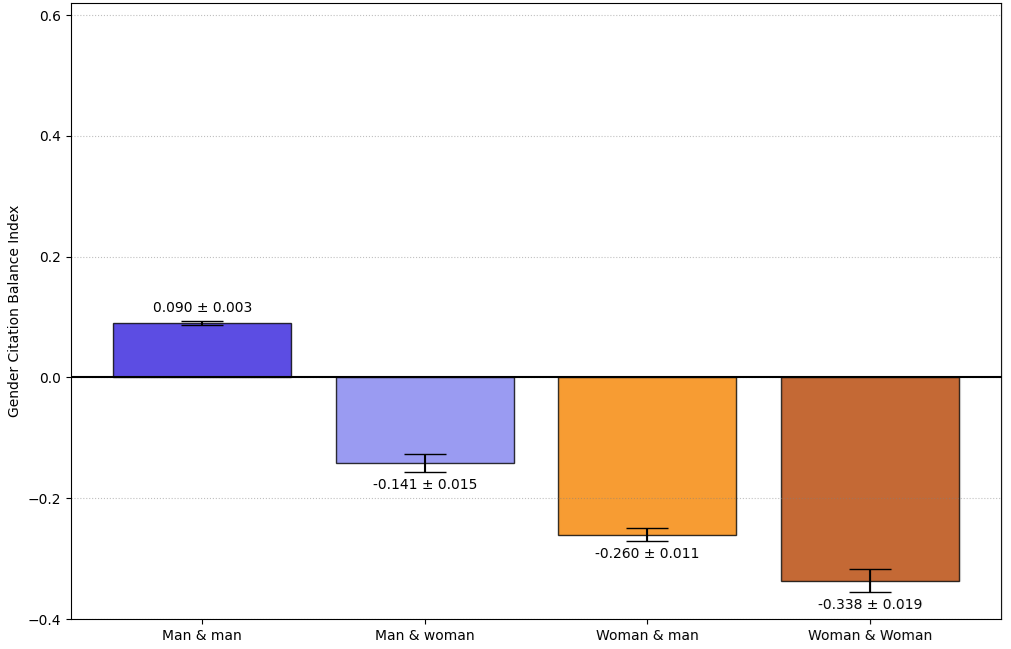}
    \caption{Citation biases among MM authors}
    \label{fig:citation_patterns_MM}
     \end{subfigure}

 \hfill
     \begin{subfigure}[b]{0.5\textwidth}
         \centering
      \includegraphics[width=0.85\linewidth]{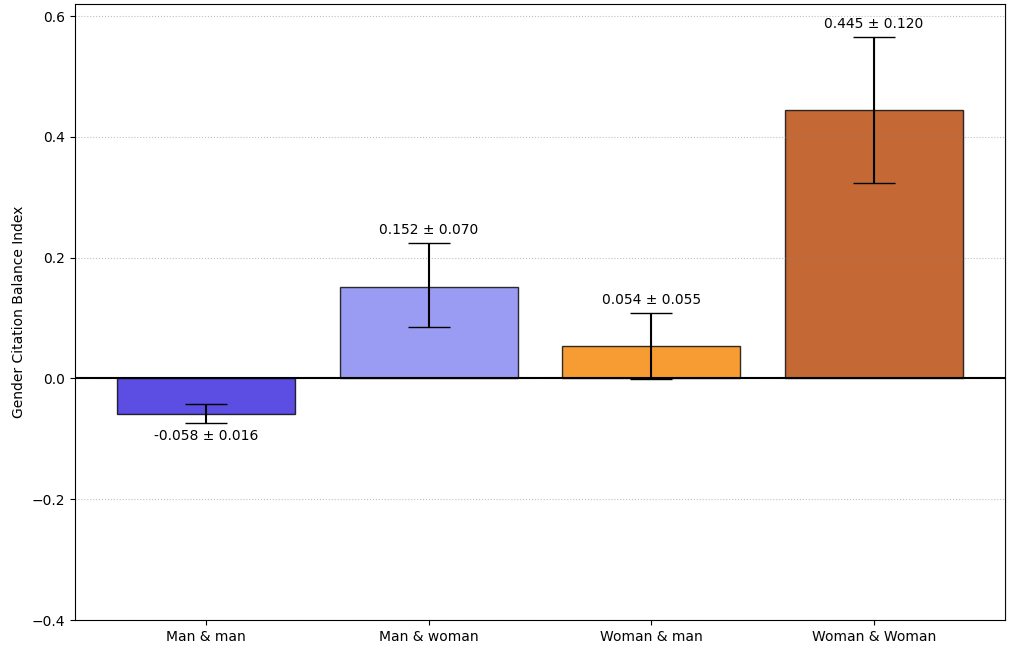}
    \caption{Citation biases among WW authors}
    \label{fig:citation_patterns_WW}
   \end{subfigure}
   
     \hfill
     \begin{subfigure}[b]{0.5\textwidth}
         \centering
    \includegraphics[width=0.85\linewidth]{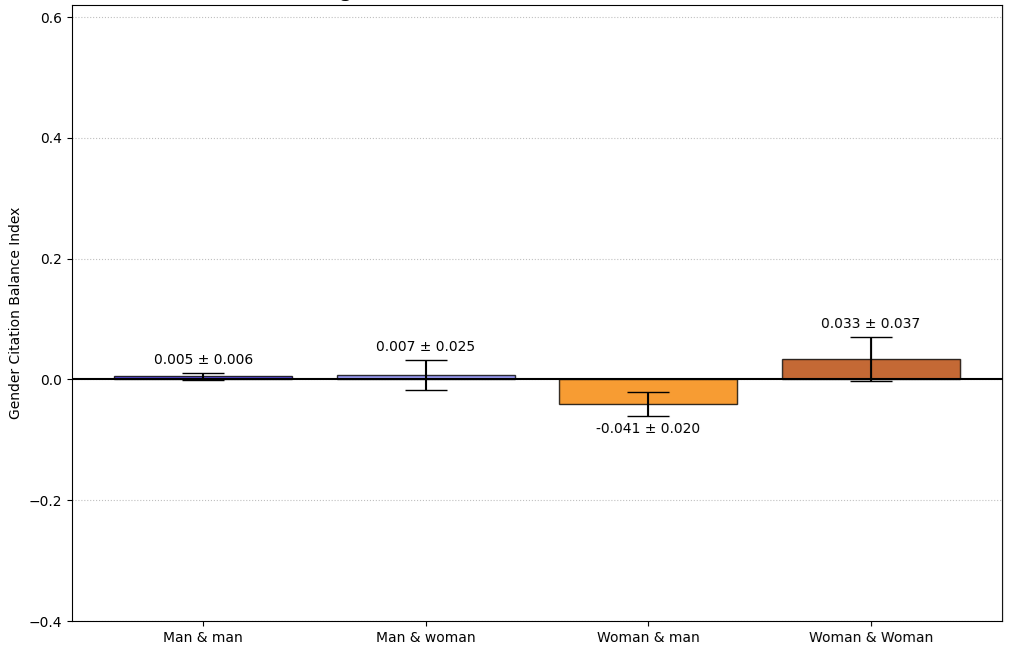}
    \caption{Citation biases among W$\cup$W authors}
    \label{fig:citation_patterns_W∪W}
     \end{subfigure}

        \caption{Citation biases in the SE literature  (2009 - 2024)}
        \label{fig:three graphs}
\end{figure}

\subsubsection{Men usually tend to cite men more than women} our results show men usually cite men more often than women. Figure \ref{fig:citation_patterns_MM} illustrates that pattern. Thus, it shows the citation behavior of the MM category toward all the analyzed gender categories. That Figure shows the MM category usually over-cites the MM category. This pattern aligns with \textit{homophily} \cite{b19}, and is in accordance with the outcomes of Tahamtan et al. \cite{b29}'s review that concludes that male authors usually prefer to cite male authors over female authors. This pattern is also in accordance with the \textit{Matthew effect} \cite{b24, b43}. The latter conveys the idea that research conducted by men is perceived as the most central and critical in a field \cite{b24, b43}. This pattern is also in accordance with the \textit{ Matilda effect} \cite{b28, b52}. The latter conveys the idea that the achievements of women are less acknowledged or research findings made by women are usually misallocated
to other male authors \cite{b43, b28}. Note that, the dominance of men in the SE journals authorship (see Figure     \ref{fig:authorship_trends}) significantly amplifies their impact on gendered citation practices. Hence, since male authors are the ones citing the most, they are the ones who drive gendered citation practices.

\subsubsection{Women usually tend to significantly cite women over men} our results also show that women tend to cite other women more that men. Figure \ref{fig:citation_patterns_WW} illustrates that pattern. More specifically, that Figure shows the citation behaviour of the WW category toward each of the gender categories under analysis. The existence of this pattern may be attributed to the solidarity efforts female authors have made over time to ensure female-led papers get cited more often, with the hope of achieving citation fairness in the field. This pattern is in accordance with the conclusions of Tahamtan et al. \cite{b29} who stated that "\textit{women are three times more likely to cite other female researchers}". This pattern is also in accordance with the conclusions of Dworkin et al. \cite{b11} who stated that women usually make conscious efforts to cite other women's papers. 

\subsubsection{Citation bias decreases when women co-author papers}
\begin{figure*}
    \centering
    \includegraphics[width=0.72\linewidth]{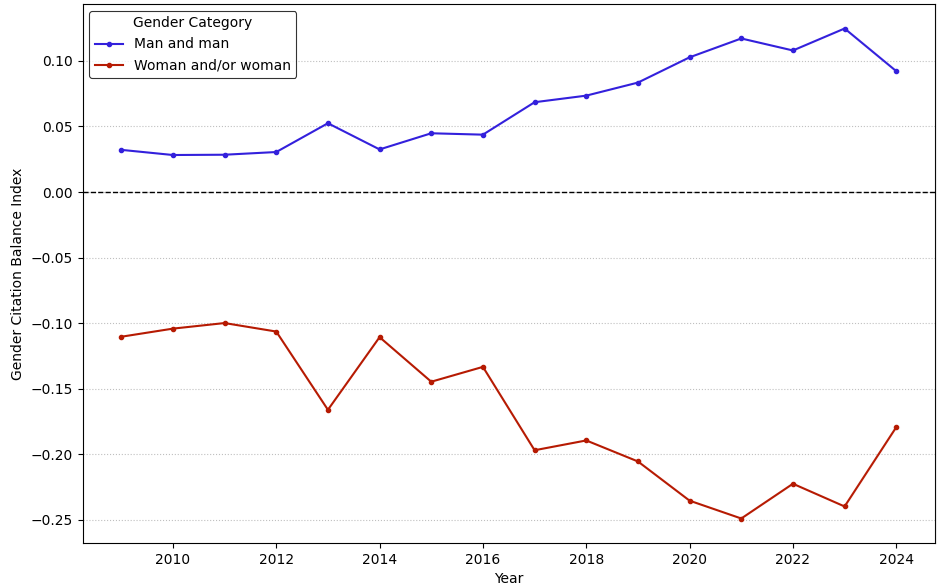}
    \caption{Temporal evolution of the Gender Citation Balance Index for the W$\cup$W category (2009 - 2024)}
    \label{fig:prelim_analysis_WuW}
\end{figure*}

\begin{figure*}
    \centering
    \includegraphics[width=0.72\linewidth]{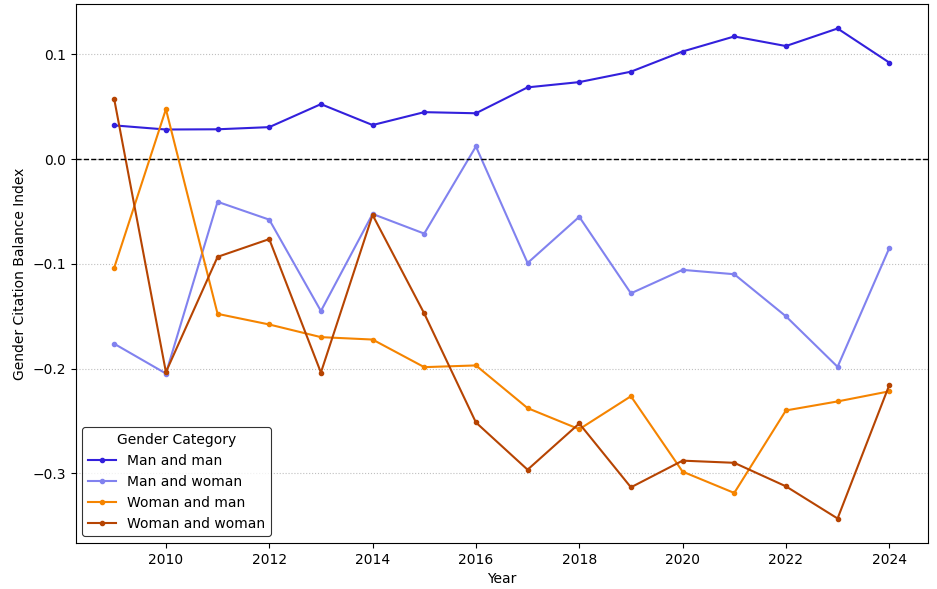}
    \caption{Temporal evolution of the Gender Citation Balance Index for all the gender categories (2009 - 2024)}
    \label{fig:prelim_analysis_all}
\end{figure*}

Interestingly, when it comes to the W$\cup$W category, the citation bias is nearly non-existent as shown on Figure \ref{fig:citation_patterns_W∪W}. The latter shows the citation behaviour of the W$\cup$W category toward each of the four gender categories under analysis. This pattern is in accordance with Dion et al. \cite{b43} who noted that the significant presence of female researchers in a field could contribute to the elimination of the citation gender gap.  Still, as Figure \ref{fig:prelim_analysis_WuW} shows, the values of the Gender Citation Balance Index for the W$\cup$W category are clearly lagging behind. Besides, as Figure \ref{fig:prelim_analysis_all} shows, the values of the Gender Citation Balance Index for the WW category remains significantly lower than the ones of the MM category. Thus, Figures \ref{fig:prelim_analysis_WuW} and \ref{fig:prelim_analysis_all} both suggest that the gendered citation bias may be getting worse through time. Noteworthy, the scarcity of women in the SE journals authorship (see Figure \ref{fig:authorship_trends}) significantly reduces the impact women have on gendered citation practices.

\smallskip
\noindent\fbox{%
    \parbox{\linewidth}{%
    \smallskip
    Our results show there are citation biases in the SE literature, female authors being the most impacted. Thus, raising awareness on this issue is key to foster fairness in citation practices.

    }}

\section{Discussion and implications}
\label{section5}
Our results are similar to the ones the literature (e.g., \cite{b5, b6, b9, b11, b14, b29, b36, b39}) obtained and/or discussed in other disciplines (e.g., neuroscience, physics,  communication). Thus, gender disparities are pervasive in various  disciplines including computing (SE) and are the fruit of longstanding systemic issues.
 Accordingly, our results indicate that some efforts  need to be done to achieve fairness in citation practices in the SE field. As stated in Section \ref{section2}, such efforts may consist in the inclusion of CDSs in manuscripts submitted for publication to SE journals and conferences.  Including CDSs in manuscripts could foster fairness in citations by promoting equal opportunities for researchers, and  balancing social justice scales
 \cite{b37}. In many fields, the literature (e.g., \cite{b4, b7, b10, b11, b14, b19, b25,  b51}) advocates for  the inclusion of CDSs in manuscripts. 
 Still, promoting diversity in reference lists should not be detrimental to the compliance with established standards of
citation ethics \cite{b25}. 

Efforts to improve fairness in citation practices may also consist in redefining the power dynamics across scientific communities,  industry, and academia to foster scientific inclusion (e.g., fair representation of scientific contributions) in the SE field. The development of fairness-centred synergies between these stakeholders could be key to improve the matter. This is notably possible by developing programs fostering international collaboration for female researchers \cite{b1}. Such collaborations would be able to tap into various perspectives to accelerate the pace of scientific and technological advances in the SE field. Such advances would inevitably yield a positive impact on the development of the society. Such international  collaborations usually yield papers that are highly cited \cite{b29}.

Nearly half the population of each country is impacted by gender disparities in science and the lack of opportunities it yields for women \cite{b1}. Citation biases, as observed in our study, exacerbate such disparities. To tackle gender disparities, it is therefore  crucial for each country to also revisit its policies to enhance women’s participation in the scientific workforce \cite{b1}. Hence, in line with Lariviere et al. \cite{b1}'s conclusions, to avoid the repetition of past order that led to gender disparities, we contend there is a need to integrate into policies the various local contexts (e.g. social, cultural, economic and political contexts) in which scientific activities are performed.
    
\section{Threats to Validity}
\label{section6}


As other existing approaches (e.g.,  \cite{b5, b6, b9, b11, b13, b14, b19, b43}), we also considered gender as a binary attribute. Hence, we did not take into account all the possible gender-related identities. This limitation stems from the capabilities of the gender assignment tools we used to determine the gender of authors.  More specifically, these tools deal with the gender as a binary attribute. This does not sufficiently reflect the gender diversity of authors as some of them may not have a binary identity. Still, finding non-binary information is a very difficult task for studies centered on fairness analysis in general \cite{b32, b33, b34, b35}, especially since such information can be difficult to infer. 
However, to better foster diversity in citation practices, it is crucial to address that limitation in future work.

As in most approaches in the literature (e.g., \cite{b5, b6, b9, b11, b47}), and in the approaches we adapted (e.g., \cite{b9, b11}) as well as their supporting tools, our citation analysis  focuses on the first and last authors of each paper in our dataset. This simplifies the citation analysis but does not allow capturing all the authorship nuances. Future work should focus on finding more efficient ways to capture authorship in citation analysis.

Our analysis only focuses on the gender of authors. This hinders the generalization of our study outcomes to other attributes and/or under-represented groups. 
Thus, there is a need to also explore other attributes such as race, prolificity, disability, class, seniority, and citizenship, as well as their intersectionality with gender \cite{b6, b11, b14}. This could yield a more robust citation analysis. 
This could also help create a broader database of
under-represented scholars \cite{b11} and help devise solutions to foster fairness in citation practices. 


\section{Conclusion and future work}
\label{section7}
The study we reported in this paper allowed concluding that the gendered citation practices adopted in the software engineering field are biased. Our results suggest the SE community needs to make efforts to fairly cite female authors.
\newline
\indent
 Our results suggest that it is crucial to encourage journal editors and conference chairs to recommend the inclusion of citation diversity statements in manuscripts submitted for publication in software engineering journals and conferences. Developing tools that will automatically generate and include citation diversity statements in such manuscripts is also key to promote fairness in citation practices. This is in accordance with Bruton et al. \cite{b37} who advocate for the development of tools and initiatives allowing to improve citation practices.
\newline
\indent
Although the results we obtained in our  study are promising, we still need to improve the proposed approach. For instance, when conducting our study, we only focused on a single  attribute that has been widely analyzed in several other disciplines: the gender. In future work, we will investigate if additional attributes could also be a source of citation bias. 
\newline
\indent
Future work will also study the impact of citation practices  on researchers' careers (eg., recruitment, promotion, leadership, awards, publication rates,  collaboration opportunities).

\section*{Full Citation Diversity Statement}
In various disciplines,  literature (e.g., \cite{b5, b6, b9, b11, b12, b20, b43})  concluded there are  citation biases toward authors belonging to some socio-cultural groups (e.g., women, Black and Hispanic). We acknowledge these biases and, in accordance with citation ethics \cite{b37}, we have made some efforts to mitigate them when creating our reference list. 
With respect to gender categories, our reference list (excluding self-citations) contains 27.45\% man/man, 17.65\% man/woman, 29.41\% woman/man, 11.76\% woman/woman and 13.73\% unknown gender authorship. With respect to race categories, our reference list (excluding self-citations) contains 64.71\% white/white, 5.88\% white/author of colour, 17.65\% author of colour/white, 5.88\% author of colour/author of colour and 5.88\% unknown race authorship. We performed gender and race classifications using respectively \textit{Gender-API} and \textit{Ethnicolr}'s Florida Voter Registration Model. These tools either deal with  gender as a binary attribute or  consider limited racial identities. Future tools should address such limitations.

\section*{Acknowledgment}
We thank the LURA (Lassonde Undergraduate Research Award)  program for funding this research.


\end{document}